\documentclass[twocolumn,showpacs,showkeys,preprintnumbers]{revtex4}
\usepackage{graphicx}
\begin{document}
\title{Modeling rhythmic patterns in the hippocampus}
\author{A.I. Lavrova}
\email{aurebours@googlemail.com}
\affiliation{Institute of Physics, Humboldt-University at Berlin,
Newtonstr. 15, 12489, Berlin, Germany}
\author{M.A. Zaks}
\email{zaks@math.hu-berlin.de}
\affiliation{Institute of Mathematics, Humboldt-University at
Berlin, Rudower Chaussee 25, 12489, Berlin, Germany}
\author{L. Schimansky-Geier}
\email{alsg@physik.hu-berlin.de}
\affiliation{Institute of Physics, Humboldt-University at Berlin,
Newtonstr.15, 12489, Berlin, Germany}

\date{\today}

\begin{abstract}
We investigate different dynamical regimes of neuronal network in
the CA3 area of the hippocampus. The proposed neuronal circuit 
includes two fast- and two slowly-spiking cells 
which are interconnected by means of dynamical synapses.
On the individual level, each neuron is modeled by FitzHugh-Nagumo equations.
Three basic rhythmic patterns are observed: gamma-rhythm in which
the fast neurons are uniformly spiking, theta-rhythm in which the
individual spikes are separated by quiet epochs, 
and theta/gamma rhythm with repeated patches of spikes.
We analyze the influence of asymmetry of synaptic strengths 
on the synchronization in the network
and demonstrate that strong asymmetry reduces the variety
of available dynamical states.
The model network exhibits multistability; this results in occurrence
of hysteresis in dependence on the conductances of individual
connections. We show that switching between different
rhythmic patterns in the network depends 
on the degree of synchronization between the slow cells.
\end{abstract}

\pacs{87.18.Sn, 87.19.ll}
\keywords{Synchronization, neuronal networks, multistability,
oscillatory regimes}
\maketitle

\section{Introduction}
Rhythmic behavior is a basic property of biological systems
and plays an important role
in a variety of physiological processes~\cite{Winfree}. In particular,
hippocampal neurons of the human brain are able to generate rhythmic
oscillations in several frequency ranges; among these, prominent roles belong
to fast (gamma-rhythm, 30$\div$65 Hz) and slow (theta-rhythm, 3$\div$10 Hz)
signals and to mixed regimes in which fast and slow oscillations
alternate. Theta-rhythm has been demonstrated to be responsible for
coding of spatial information \cite{Keef}, synaptic modification
of intrahippocampal pathways, as well as for assembling and
segregation of neuronal groups \cite{Bur}. In its turn, oscillatory
activity in the gamma-frequency band is involved in information
transmission and storage \cite{Harr}. Isolated parts (so-called
CA1, CA3, DG etc.) of hippocampal formation where these
regimes are observed,  have been subjected to numerous
experiments \cite{Haj, Glov, Gloveli, Macca, Mann, Klaus}. In
particular, it has been established that the neuronal network in the CA3 region
of the hippocampus can exhibit three qualitatively different oscillatory
states (theta-, gamma- and the mixed ``theta/gamma'') and is able to
dynamically switch between them \cite{Gloveli}.
Transitions between different regimes in brain neuronal networks can be
provoked not only by variation of parameters, but by variation of initial
conditions as well~\cite{Fro,Durst,Kass,Sasaki}; the latter may occur due to
changes in the baseline potential~\cite{Kass} or to perturbations
of initial ionic concentrations~\cite{Hahn}.

Cells, which constitute a hippocampal network, differ in their morphological,
electrophysiological and neurochemical properties \cite{Ander}.
Synaptic interactions between several types of cells: the pyramidal ones,
basket, septum and  oriens-lacunosum-moleculare (OLM) cells are responsible
for the generation of different rhythms~\cite{Klaus}.

Besides experimental studies, transitions between various rhythms
and multistability phenomena in isolated cells as well as in neuronal
networks have been subjected to theoretical and numerical
modeling~\cite{Gloveli,Kopell,Siek,Frohl,Rab1}. Typically, such
models involve various cell types as well as large numbers of
ionic currents and compartments. In the detailed models~\cite{Traub,Frohl}
as well as in simpler biophysical approximations~\cite{Kopell}
the cells, responsible for the generation of rhythmic patterns,
are typically described by equations of the Hodgkin-Huxley type.

Role of intercellular synchronization in the transitional
processes is an object of ongoing discussions. In particular, it
has been shown~\cite{Sing} that a switching between two states of
the system can forecast desynchronization in a bursting activity
of the hippocampus. The strength of intercellular coupling governed by
synaptic conductances is apparently a crucial factor in this
context~\cite{Gloveli,Rab}. For example, studies of detailed
thalamocortical model~\cite{Rab1} have shown that presence or absence
of bistability in the neuronal circuit can depend on parameters
of the electrical coupling
between neurons as well as on the synaptic conductance.

High degree of complexity, obligatory in realistic models,
hampers revealing of principal mechanisms and control parameters
of transitions between different types of oscillations.
Therefore, along with accurate reproduction of experimental observations,
one of the main aims of the modeling remains a consideration
of general dynamics in the appropriate class of circuits,
as well as an understanding of the underlying dynamical mechanisms
which are responsible for different rhythmic patterns.
This raises a question
about a minimal module of the hippocampal network: a small set of cells
capable of reproducing all basic typs of experimentally observable
rhythms~\cite{Kopell}.
Embedded into intricate networks of neurons, such modules can play a role
of pacemakers which ensure persistence of optimal oscillatory states in
the entire ensemble.

Below we discuss a minimalistic dynamical model of the hippocampal circuit
in the area CA3, which is able to produce several rhythmic patterns.
For this purpose we simplify the basic module of the circuit in
such a way that maintenance of the essential types of dynamical
regimes would be controlled by a concise number of principal
parameters. Since we do not aim at detailed reproduction of
intrinsic cellular dynamics but restrict ourselves to rhythmic
aspects, we reduce the description of each element to the
classical simplified model of neuronal oscillations: the
FitzHugh-Nagumo (FHN) equations. Simplification of elements in this case
does not imply trivialization of dynamics:
ensembles of the FHN oscillators are known to exhibit
different kinds of collective behavior which, depending on the
topology and strength of interaction, range from subthreshold oscillations
via mixed-mode intermittency to tonic spiking~\cite{zssgn}. 
Interplay between connectivity and coupling strength in networks of
heterogeneous FHN oscillators can induce or destroy synchrony~\cite{hennig}.
Working with FHN neurons renders the well-understood dynamics on the
level of individual elements and allows us to concentrate
on the governing role of interactions between them.

In Sect.~\ref{sect_model} we introduce the model equations and discuss the
choice of appropriate parameter values. Sect.~\ref{sect_results} begins
with characterization of the principal rhythmic patterns;
in particular, the role of the phase shift between the OLM cells
is discussed. Further we proceed to the description of multistability
and transitions between the patterns.
We demonstrate that the model reproduces the basic rhythms
observed experimentally, as well as hysteresis between them.
Finally, we show how the symmetry or asymmetry in the arrangement
of synaptic connections influences the existence
and properties of dynamical states.

\section{Model equations}
\label{sect_model}

In a recent paper \cite{Gloveli}, a kind of minimalistic network for
studies of different regimes in the hippocampal area CA3 has been proposed.
This network includes main types of hippocampal cells and consists of
five elements: two fast-oscillating cells (so-called basket cells),
two slow cells (OLM -- oriens/lacunosum-moleculare associated -- cells)
and one two-compartmental pyramidal cell.
The cells are synaptically connected in all-to-all topology, with one
exception: there is no direct connection between the slowly spiking cells.
Accordingly, the pyramidal cell activates the rest of the cells which,
in turn, inhibit each other and the pyramidal cell.
The cells are described within the framework of the Hodgkin-Huxley formalism.
After taking into account all relevant currents as well as
kinetics of synaptic variables, the resulting system
constitutes a set of 41 ordinary differential equations.
Notably, the choice of coupling coefficients
employed in~\cite{Gloveli} ensured a certain degree of asymmetry in the network:
in particular, each basket cell was coupled to both OLM-cells through
connections with different conductivity.

Numerical simulations of that model have confirmed that
transitions between different regimes depend on the
strength of coupling between the cells.
Typically, a switching between different rhythms is preceded by onset
of a certain phase shift between two slow-spiking cells.
Within the model of~\cite{Gloveli}, this shift can be ensured e.g. by
a judicious choice of appropriate initial conditions.

\begin{figure}[h]
\includegraphics[width=0.39\textwidth]{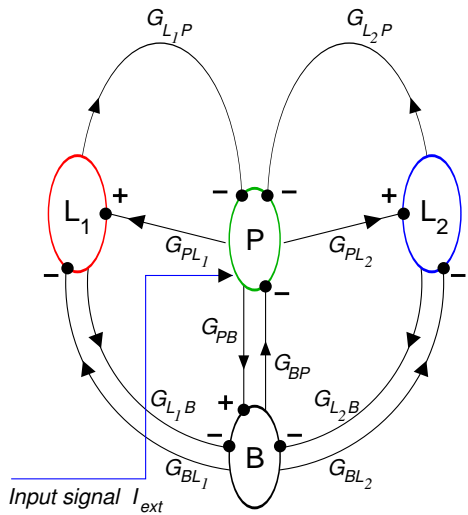}
\caption{(Color Online). Sketch of the cellular network.
$L_{1,2}$: slow OLM cells; $P$: fast pyramidal cell;
$B$: basket cell.
Arrows: directions of currents. Filled circles: synapses.
Pluses: excitatory synapses; minuses: inhibitory synapses.
Synaptic connections are
characterized by conductances $G_{ji}$.
External input $I_{ext}$ affects only the pyramidal cell $P$.}
\label{network_sketch}
\end{figure}

In the present work we further reduce the complexity of the model:
we decrease the number of
cells and replace the Hodgkin-Huxley setup by the FitzHugh-Nagumo equations.
The resulting  network, sketched in Fig.\ref{network_sketch},
includes one-compartmental pyramidal cell $P$,
one basket cell $B$ and two OLM-cells $L_1$ and $L_2$.
All cells are mutually synaptically connected; the only exception are
two slow  OLM-cells which do not communicate directly.
Synaptic connections are governed by the synaptic
conductances $G_{ji}$ with $j,i=L_1,\,L_2,\,P,\,B$ (and $G_{jj}$=0).
Leaving in the network only one basket cell $B$, we take into account, however,
inequality of the cross- and direct- connections 
as a potential source of a phase shift
between the slowly spiking cells. For this purpose, it is sufficient to allow for
asymmetric synaptic inputs from the basket cell to $L_1$ and $L_2$:
in general, $G_{BL_1}\neq G_{BL_2}$.
The rest of connections is set symmetrically:
$G_{PL_1}$=$G_{PL_2}$=$G_{PL}$, $G_{L_1P}$=$G_{L_2P}$=$G_{LP}$,
$G_{L_1B}$=$G_{L_2B}$=$G_{LB}$.

Within this simplified description, it appears reasonable to keep only one
activating input which acts upon the fast-oscillating cell $P$. The latter
excites all other cells, which inhibit each other as well as $P$ itself.

We use below the form of equations in which all variables,
and, hence, all parameters are measured in dimensionless units.
Each cell of the network obeys the standard set of coupled
FitzHugh-Nagumo equations:
\begin{eqnarray}
\frac{dv_i}{dt}&=&v_i -\frac{v_i^3}{3}-u_i+I_{\rm ext}\delta_{i,P}
+\sum_j I_{\rm syn}^{(ji)}, \nonumber\\
\label{equations}\\[-3ex]
\frac{du_i}{dt}&=& \varepsilon_i\,(v_i+a-b\,u_i).\nonumber
\end{eqnarray}
where $v_i$ is the membrane potential of the $i$-th cell
(recall that $i=L_1,L_2,P,B$), and $u_i$ is the respective
membrane variable.
The values of parameters $a$ and $b$ are the same for all cells.
The Kronecker-delta $\delta_{i,P}$ ensures that only the pyramidal cell
is externally excited by the current $I_{\rm ext}$.

The synaptic input $I_{\rm syn}^{(ji)}$ in Eqs (\ref{equations}) -- the
current from cell $j$ to cell $i$ -- is defined as
$$I_{\rm syn}^{(ji)}=G_{ji}s_{ji}\,(E_{\rm ex,in}-v_i)$$
and is governed by kinetic equation for the synaptic variable $s_{ji}(t)$
\cite{Kopell}:
\begin{equation}
\frac{ds_{ji}}{dt}=
\frac{A}{2}\left(1+\tanh\frac{v_j}{v_{vsl}}\right)(1-s_{ji})-Bs_{ji},
\label{sin_var}
\end{equation}

For simplicity, we assume that all synapses obey quantitatively the same
kinetics. This is achieved by using the same values of $A$, $B$ and $v_{vsl}$
in (\ref{sin_var}) for all $s_{ji}$.
Hence, evolution of $s_{ji}$ depends (via $v_j$) only on $j$,
and the number of independent synaptic variables equals
the number of cells in the network.
Altogether, the dynamical system includes 12 ordinary differential equations:
8 equations for the FitzHugh-Nagumo variables along with
4 equations for the synaptic variables.

Originally, the FitzHugh-Nagumo equations were written as
a crude simplistic model which mimics the characteristic features of the Hodgkin-Huxley dynamics. Hence, their parameters do not allow for
direct biological interpretation and serve the mere
purpose of supporting the proper dynamical regimes.
To illustrate possible kinds of dynamics in the considered network, we fix
the parameters of the FitzHugh-Nagumo equations (\ref{equations})
at the values which ensure that in absence of external currents
every isolated cell is at rest;
if perturbed by sufficient external input, it displays full-scale
oscillations. The values $a$=0.5, $b$=0.8 which we use for our calculations,
are close to those employed by FitzHugh~\cite{FH_69}.

The parameter $\varepsilon_i$ in the FitzHugh-Nagumo
equations (\ref{equations}) determines the timescale of
oscillations in the isolated $i$-th cell: in the leading order,
period of relaxation oscillations in a cell is inversely proportional
to $\varepsilon_i$.  The dimensionless number $\varepsilon_i$
can be interpreted as ratio of two {\em dimensional} timescales of the
neuron: that of the fastest relaxation to that of the slowest one.
In the case of the pyramidal and the basket cells these are the relaxation
times of, respectively, potassium and sodium currents; the values of
$\varepsilon_i$ for these two cells can be viewed as (approximately) the same.
In contrast, in the OLM-cells the slowest processes
involve relaxation of specific so-called
hyperpolarization-activated currents~\cite{Macca,Gil},
and their values of $\varepsilon_i$ should be
chosen substantially lower. Accordingly,
we set $\varepsilon$=0.3 for  the fast basket and pyramidal cells,
whereas for slow $L$-cells the value $\varepsilon$=0.04 is adopted.

Equations  (\ref{sin_var}) for synaptic variables are phenomenological
as well; the values of parameters  $A$, $B$ and $v_{vsl}$ are commonly
obtained by fitting the experimentally measured dependencies.
By setting $A$=1, $B$=0.3 and $v_{vsl}$=0.1 we model the typical
situation in which opening of channels occurs faster -- but not much faster --
than their subsequent closure.

The value of $E$ in the expression for synaptic current depends on
the kind of connection.
For all excitatory synapses we put $E_{\rm ex}=0$, whereas for inhibitory
synapses $E_{\rm in}$=--5 is set. The latter value ensures that the
difference $E_{\rm in}-v_i$, and thereby the corresponding synaptic current
$I_{\rm syn}^{(ji)}$ stay negative at all times,
and the connection is indeed inhibitory.

Concerning synaptic conductances $G_{ji}$, we choose the values
which, dimensionalized in units of 0.01$Ohm^{-1}m^{-2}$,
have the order of magnitude of those, reported in~\cite{Gloveli}.
In a sense, conductances predetermine the
relative importance of individual cells in the ensemble.
Since external current acts only upon the pyramidal cell,
synaptic connections from this cell to the rest of the module
should be strong enough in order to activate the otherwise silent basket and
OLM-cells;
this is ensured by assigning to them relatively high values $G_{PL}=0.7$
and $G_{PB}=0.57$. A lower conductivity is assigned to the backward
connection from the basket to the pyramidal cell: $G_{BP}$=0.1.
Further, we introduce {\em asymmetry} between the left and right halves of
the network by fixing different values for connections
from the basket to the OLM-cells:
$G_{BL_1}$=0.06 in contrast to $G_{BL_2}$=0.03.

The slow OLM-cells are driven by the pyramidal cell; intensity of
their inhibitory feedback to the driving element depends on conductances
$G_{LP}$ and (mediated by the basket cell) on conductances $G_{LB}$.
If both $G_{LB}$ and $G_{LP}$ are set to zero, the feedback is absent,
and the OLM-cells play the passive role. Each of them oscillates periodically:
it is phase-locked to the oscillations of the driving fast subsystem formed
by the pyramidal cell and the basket cell.
However, due to asymmetry, the slow cells can be locked
(and under the quoted parameter values are indeed locked)
to the fast subsystem in different locking ratios. Asymmetry of rhythmic patterns
persists for sufficiently low values of $G_{LB}$ and $G_{LP}$: this is visualized
in the lower panel
of Fig.\ref{gamma_1} where the slow cells oscillate in the ratio 2:3.
Increase of conductances $G_{LB}$ and/or $G_{LP}$ intensifies the feedback;
in spite of the absence of the direct connection, the OLM-cells interact
through the mediation of the fast cells, and their periods get adjusted.

In order to concentrate on the role of connection from the OLM-cells
to the pyramidal cell, we use throughout  this work the conductance
$G_{LP}$ as a controlling parameter and study transitions which occur
in the course of its variation. As for the
connection from the OLM to the basket cell, it should be weak enough
in order not to damp the module dynamics, hence we fix for our calculations
(with the only exception of the plot in Fig.\ref{gamma_1})
the value $G_{LB}=0.01$.  This value is sufficiently high
to ensure that the feedback via the basket cell synchronizes the slow cells
in the frequency ratio 1:1, even in the absence of direct feedback through
the $LP$-connections.

The steady solution of equations corresponds to the quiescent non-spiking state.
Intensity of the constant external input $I_{\rm ext}$
(recall:  it affects only the $P$-cell)
should ensure destabilization of the equilibrium
and sustainment of oscillations in the system.
Under $G_{LP}$=0.037 and the aforementioned fixed values of other parameters,
the equilibrium is unstable for $0.02<I_{\rm ext}<1.1$.
In the endpoints of this interval, subcritical Hopf bifurcations take place.
Location of the endpoints is almost insensitive to variation of $G_{LP}$.
Accordingly, for our numerical simulations we take the value of input from
this range: $I_{\rm ext}=0.43$.

Numerical solutions have been obtained using the software
packages XPPAUT and MATCONT.

\section{Results}
\label{sect_results}

In order to characterize different attractors of the system
and to understand possible mechanisms of switching between the regimes,
we have studied the response of the model to variation of the strength of $LP$
connection. Starting from the situation when this connection is absent,
we describe below the states which are observed in the course
of increase of $G_{LP}$.

Among the solutions, we single out the three main qualitative types of oscillations.
The difference between them lies in the behavior of the voltage
variables of fast (pyramidal or basket) neurons: in the gamma-rhythm the
voltage exhibits fast regular oscillations, whose amplitude is weakly
modulated due to interaction with slow counterparts. In the theta/gamma rhythm
certain spikes in the pattern ``fall out''; the plot
shows regular patches of spikes interrupted by nearly quiescent epochs.
Typically, theta/gamma oscillations
are periodic. Finally, the theta pattern consists
of repeated solitary spikes on the nearly quiescent background.
For a  slow OLM-cell the difference in the temporal dynamics between
three rhythms is not so pronounced; here, transitions manifest themselves
in the onset or the disappearance of a phase shift between two such cells.

\subsection{Gamma-rhythm}
\label{sect_gamma}
\begin{figure}[h]
\includegraphics[width=0.9\columnwidth]{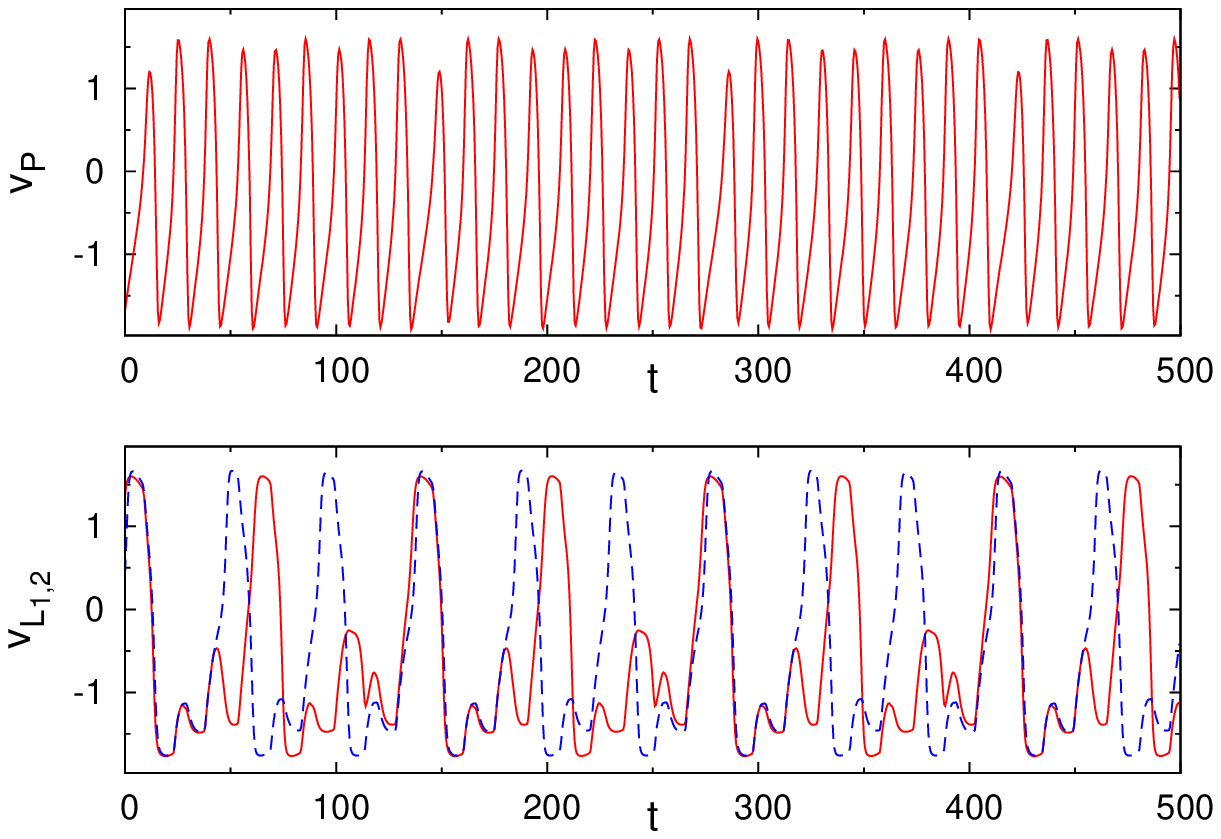}
\caption {(Color Online). Gamma-rhythm at $G_{LP}$=0.032, $G_{LB}$=0.001.
Top: pyramidal cell; periodic pattern which consists of 9 spikes.
Bottom row: OLM cells; solid curve: $L_1$; dashed curve:
$L_2$. OLM cells are locked in the frequency ratio 2:3.}
\label{gamma_1}
\end{figure}

If the value of conductance  $G_{LP}$ is set to zero,
the OLM-cells exert no inhibiting action upon the pyramidal cell;
this corresponds to removal of the upper connecting arcs in the scheme
of Fig.\ref{network_sketch}. In this situation,
the system performs periodic oscillations of the gamma type.
Consecutive spikes in the fast cells occur at close
time intervals, but the exact repetition (in other words, closure
of the limit cycle in the phase space of the system) requires several spikes.
In particular, smallness of $G_{LB}$ can result in asynchrony between
the slow cells (Fig.~\ref{gamma_1}).

\begin{figure}[h]
\includegraphics[width=0.9\columnwidth]{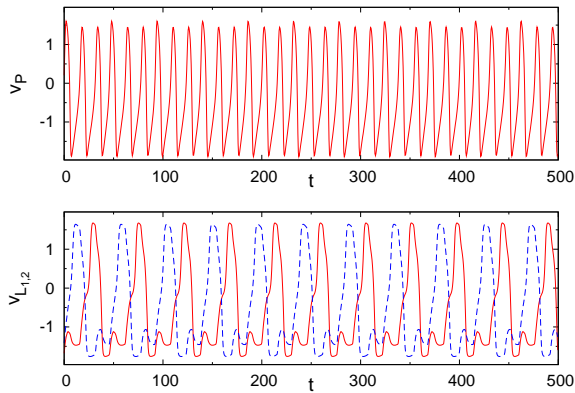}
\caption {(Color Online). Gamma-rhythm at $G_{LP}$=0.035, $G_{LB}$=0.01.
Top: pyramidal cell. Bottom row: OLM cells; solid curve: $L_1$;
dashed curve: $L_2$. OLM-cells are locked in 1:1 ratio; note phase shift
between them.}
\label{gamma_2}
\end{figure}

By continuity, rhythmic pattern of the gamma type
persists at  sufficiently low non-zero values of $G_{LP}$
as well.
An example is shown in Fig.\ref{gamma_2}.
At  $G_{LP}=0 .035$ the slow cells oscillate in the ratio $1:1$
(bottom panel), but their maxima do not occur simultaneously. In this context
it is natural to speak about the
phase shift. We introduce the phase phenomenologically: for each cell
the phase increment of $2\pi$ is assigned to every interval between two
consecutive positive maxima of voltage; between the maxima, the phase is linearly
interpolated. The phase shift between oscillations shown in the bottom panel of
Fig.~\ref{gamma_2} is close to $2\pi/3$.

\subsection{Theta/gamma rhythm}

\begin{figure}[h]
\includegraphics[width=0.9\columnwidth]{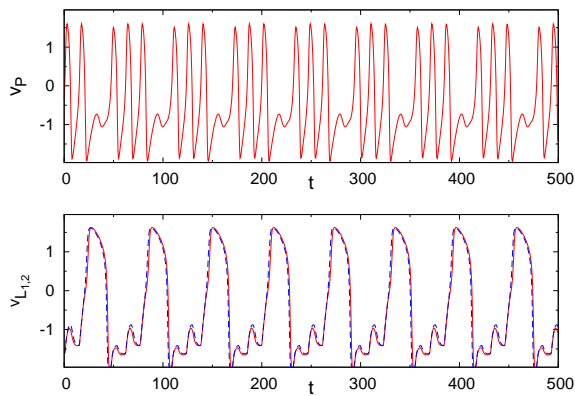}
\caption {(Color Online). Theta/gamma-rhythm close to its onset;
$G_{LP}$=0.05.
Top: pyramidal cell.
Bottom: OLM cells; solid curve: $L_1$; dashed curve: $L_2$.}
\label{theta_gamma}
\end{figure}

In a broad range of values $G_{LP}$ above 0.0362 an appropriate choice of initial
condition results in a different rhythmic pattern:
the system exhibits theta/gamma oscillations.
The bottom panel of Fig.~\ref{theta_gamma} indicates that in the
theta/gamma state both slow cells oscillate synchronously;
there is no phase shift between them.

\subsection{Theta-rhythm}

\begin{figure}[h]
\includegraphics[width=0.9\columnwidth]{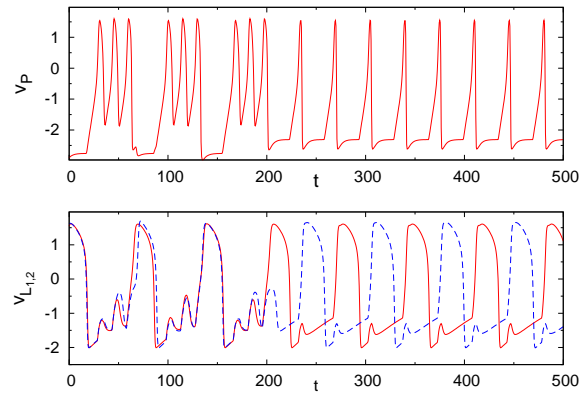}
\caption {(Color Online). Breakdown of theta/gamma rhythm and onset of theta rhythm,
$G_{LP}$=2.274.
Top: pyramidal cell. Bottom: OLM cells; solid curve: $L_1$;
dashed curve: $L_2$.
Note in-phase oscillations before the switch and antiphase immediately
after it.}
\label{switch}
\end{figure}

At sufficiently high values of  $G_{LP}$ theta/gamma rhythm is replaced
by theta-oscillations. Switching between the two regimes is visualized
in Fig.~\ref{switch}: the motion starts as a
typical theta/gamma oscillation, with patches of
spikes of the fast cell; in the course of time evolution, at around
$t\approx 200$, it abruptly changes the pattern 
and acquires the characteristic shape of
theta rhythm with equidistant solitary spikes of the fast cell.
Remarkably,  the slow cells, synchronous on the initial
stage, develop during the switch the phase shift of $\pi$: 
in the theta-rhythm the OLM cells oscillate in antiphase.

\subsection{Hysteresis}

Over large intervals of values of  $G_{LP}$, different rhythmic patterns
coexist as attractors of Eqs (\ref{equations},\ref{sin_var}): at low values
of $G_{LP}$, depending on the initial values of the variables, the system
can exhibit gamma- as well as theta/gamma-oscillations, whereas at moderate
and high values of $G_{LP}$ there is a hysteresis between
theta/gamma- and pure theta- patterns.

\begin{figure}[h]
\includegraphics[width=0.99\columnwidth]{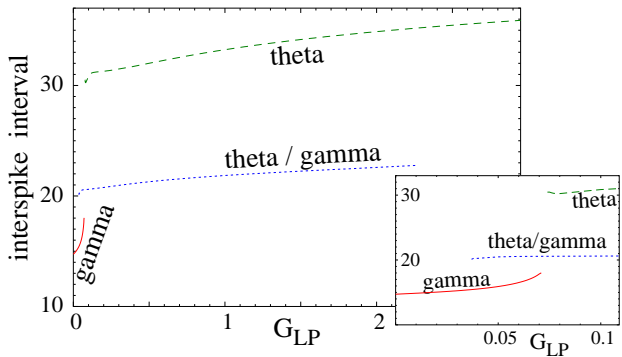}
\caption{(Color Online). Duration of interspike intervals in dependence on $G_{LP}$
for different rhythmic patterns.
Solid line: gamma-rhythm. Dotted line: theta/gamma pattern.
Dashed line: theta rhythm.
Inlet: range of low values of the conductance $G_{LP}$. }
\label{hyst}
\end{figure}

For slow OLM cells, neither a transition between gamma- and theta/gamma-states,
nor the further growth of $G_{LP}$ significantly affect the period
of their oscillations. In contrast,
periodicity in spiking patterns of the fast cells $P$ and $B$
changes quite noticeably. A convenient characteristics is delivered
by the average duration of interspike interval   (ISI):
the mean distance between the positive maxima of voltage in a cell.
In Fig.~\ref{hyst}, we plot this characteristics
for membrane potential oscillations in the
pyramidal cell. The value of ISI increases from 14.73
in the gamma state at $G_{LP}=0$ to 35.94 in the theta state at $G_{LP}=3$.
Similar situation is observed for the basket cell.

\begin{figure}[h]
\includegraphics[width=0.75\columnwidth]{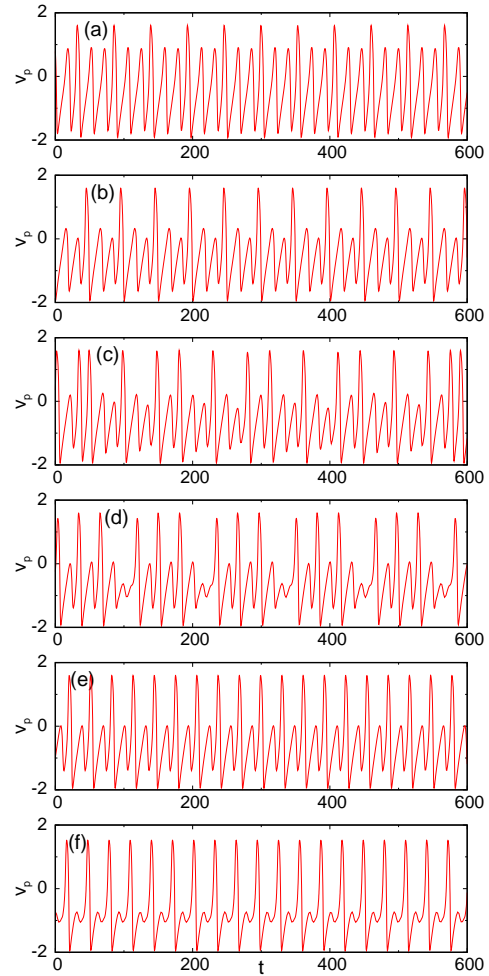}
\caption{(Color Online). Transformation of gamma- into the theta-rhythm
for the membrane voltage of the pyramidal cell.
(a) $G_{LP}$=0.07; (b) $G_{LP}$=0.0724; (c) $G_{LP}$=0.0726;
(d) $G_{LP}$=0.0732; (e) $G_{LP}$=0.0734; (f) $G_{LP}$=0.1000.}
\label{transformation}
\end{figure}

\begin{figure}[h]
\includegraphics[width=0.6\columnwidth]{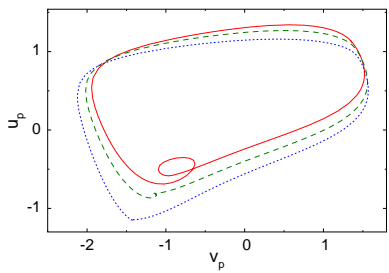}
\caption{(Color Online). Evolution of attractor shape in the phase space.
Theta-rhythm. Projection on the variables of the pyramidal cell.
Minor loops, which correspond to subthreshold oscillations, shrink,
turn into cusps and disappear.\\
Solid line: $G_{LP}$=0.09, dashed line: $G_{LP}$=0.2,
dotted line: $G_{LP}$=0.4.}
\label{cusps}
\end{figure}

A closer look shows that gamma- and theta- branches are connected;
in fact, they belong to the same continuous family of solutions.
A transformation from the gamma- to  theta-rhythm occurs
through the intermediate regime which is observed
within the small parameter range
$0.0724<G_{LP}<0.0734$. Main stages of this process,
in terms of the voltage variable of the pyramidal cell, are
presented in Fig.~\ref{transformation}.
The transformation is preceded by a gradual deformation
of the gamma-pattern: of the three initially nearly equal spikes
(cf. Fig.~\ref{gamma_2}), two spikes get diminished
(Fig.~\ref{transformation}(a,b)).
As a result, the periodic gamma-state
acquires the characteristic shape of mixed-mode
oscillations~\cite{Chaos_focus}
with alternating small subthreshold - and large-scale spiking
maxima.  At $G_{LP}=0.0726$ these oscillations form the complicated (yet
periodic, with period 541.1 !) temporal pattern,
visualized in Fig.~\ref{transformation}(c).
A small increase of $G_{LP}$ leads to partial flattening of subthreshold epochs
(Fig.~\ref{transformation}(d)); the next small increment results in the regular
alteration of a spike and a single subthreshold oscillation
(Fig.~\ref{transformation}(e)).
In fact, this state is already very close to the theta-pattern;
further increase of  $G_{LP}$ brings only quantitative changes:
the subthreshold oscillations are gradually flattened,
eliminated, and the theta state is established.
On the phase portrait (cf. Fig. \ref{cusps}),
the epochs of subthreshold
oscillations are represented by minor loops of the trajectory.
In the course of flattening, these loops get  smaller, turn into cusps and
disappear, so that only the large loops (corresponding to solitary spikes
of the theta-rhythm) persist.

Remarkably, the described transformation changes the phase difference
between the slow cells: as already mentioned, in the gamma-state
the difference is close to $2\pi/3$, and the proximity to
this value holds until the onset of mixed-mode oscillations
in Fig.~\ref{transformation}(c).
As soon as the simple temporal pattern is regained
(Fig.~\ref{transformation}(e)) the phase difference acquires
the value close to $\pi$, and the antiphase oscillations of slow
cells remain the hallmark of the  theta-state
everywhere in the range of its existence.

A similar process of deformation occurs on the other branch of solutions,
which corresponds to the theta/gamma state.
In this state, patches of full-scale spikes are separated by
subthreshold oscillations (cf. Fig.\ref{theta_gamma}); the latter
correspond to minor loops on the projections of respective
phase portraits. In the course of increase of conductance $G_{LP}$,
the shape of the attracting orbit is subjected to continuous deformation
(similarly to that shown in Fig. \ref{cusps} for theta-rhythm):
the minor loop gradually gets smaller,
shrinks into a cusp and disappears completely. This process leads to gradual
decrease and elimination of one of the maxima in the temporal pattern
of the oscillations.

As seen in Fig.~\ref{hyst}, the branch of theta/gamma oscillations is
isolated in the parameter space.
A closer look shows that its birth at low values of $G_{LP}$
begins with a very short segment of mixed-mode oscillations; they are
present close to $G_{LP}$=0.0362, but already at $G_{LP}$=0.0363
the characteristic shape of periodic theta/gamma oscillations is established.
On the other end of the branch, at $G_{LP}$=2.274, the limit cycle which
corresponds to the theta/gamma
rhythmic pattern, disappears in the saddle-node bifurcation.

\subsection{Attraction basins of coexisting states}

In further numerical simulations we studied the sensitivity
of the system response with respect to a variation of initial conditions.
At low (below 0.036) values of conductance $G_{LP}$
the oscillatory state of the gamma-type was reached for
all tested initial conditions. On the other end, at sufficiently
large (above 2.28) values of  $G_{LP}$ the system
appears to possess the unique attractor as well: now this
is the limit cycle which corresponds to the theta regime.
In between, the system is bistable.

For a case study we take the value  $G_{LP}=0.8$,
well inside the range of this parameter where the
distinct oscillations of the theta type coexist with the theta/gamma
rhythmic pattern.

The system has been scanned on a grid of initial conditions for
variables $v_i$ and $u_i$ from -2 to 2: these ranges correspond to the
maximal span of gamma and theta/gamma oscillations for pyramidal cells.
The initial size of the mesh has been taken as 0.01$\times $0.01, and has
been refined whenever necessary in order to resolve the fine details.
It turned out that the choice of eventual attractor depends mostly on
initial values of membrane potential and membrane variable of the slow cells.

\begin{figure}[h]
\includegraphics[width=0.99\columnwidth]{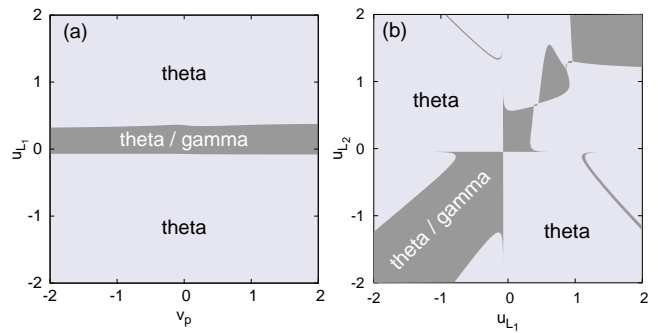}\quad
\caption{(Color Online). Projections of attraction basins of coexisting rhythmic patterns.
$G_{BL_1}=0.06, G_{BL_2}=0.03$, $G_{LP}=0.8$.
Dark grey: attraction basin of the theta/gamma rhythm. Light grey:
attraction basin of theta-oscillations.
(a): initial values for voltage of the pyramidal cell vs. membrane variable
of the left slow cell; the rest of variables is set to zero.
(b): initial values for membrane variables of two slow cells;
the rest of variables is set to zero.}
\label{initial_values_asym}
\end{figure}

To illustrate graphically the shape of the boundary between
the attraction basins in the
12-dimensional space of initial values, we choose two characteristic
intersections of this boundary with coordinate planes.
The first one (left panel of Fig.\ref{initial_values_asym}) is the plane upon
which all coordinates except  $v_p$ (voltage variable of the pyramidal cell)
and $u_{L_1}$ (gating variable of the left slow cell)  are set to zero.
The plot shows two distinct regions of theta oscillations separated by
a strip which belongs to the basin of the theta/gamma state. Both
borderlines display rather mild variations. Notably, the initial
value of $u_{L_1}$ should be sufficiently large in order to
excite the theta oscillation: initial strong contrast between the $OLM$-cells
(recall that $u_{L_2}$=0 upon this plane) would facilitate the onset of the
large phase difference between them, imminent for the theta state.

The second coordinate plane corresponds to the case when only the voltage
variables of the $OLM$-cells are initially excited whereas the rest of the
variables starts from zero values. In this projection
(right panel of Fig.\ref{initial_values_asym}), the shape of attraction
basins is much more intricate, with characteristic cusps, islands
and narrow bridges between them. Here, again, we see that proximity between
the initial states of the slow cells results, as a rule, in the eventual
onset of the theta/gamma rhythmic pattern.

Experimentally relevant is the location of the border between
the basins in terms of the membrane voltage of slow cells.
Numerics shows that this border (not presented graphically) looks quite simple:
within the studied range of $v_p$ there exists a threshold
value of voltage $v_{L_{1,2}}\approx 0.14$ below which the unit oscillates
in the theta/gamma rhythm whereas above it the theta rhythm has been
registered.

\subsection{Variation of asymmetry between BL-synaptic connections}
\label{sect_symmetry}
In order to elucidate the possible role of asymmetry in the pattern
of synaptic connections, we performed calculations at different
values of the ratio $G_{BL_1}/G_{BL_2}$. Decreasing this ratio from 2 to 1,
we have detected  qualitative distinction only at very low values of 
conductance $G_ {LP}$; otherwise, the results were largely similar
to the ones described in the previous subsections.

In the symmetric case  $G_{BL_1}=G_{BL_2}$,
the system possesses the invariant ``diagonal'' subspace
in which the values of variables for the two slow cells coincide
(hence, the phase shift between them is absent). Surprisingly, the states
from this subspace are stable not only at moderate values of $ G_{LP}$
which correspond to the attractor of the theta/gamma type,
but at low values of this parameter as well, where the temporal pattern
belongs to the gamma-type.
In contrast to the previously described gamma-oscillations, in this state
the phase shift between the slow cells is absent;
they oscillate in unison. Below we call this regime a ``symmetric
gamma-state''.
If the conductance $G_{LP}$ is set below $0.0016$,
symmetric gamma-oscillations are chaotic.
Above this value of $G_{LP}$ the periodic gamma-state is observed;
in a narrow interval close to $ G_{LP}$=0.036 its regular pattern is
transformed (via the mixed-mode) into the periodic theta/gamma oscillation.
Outside the diagonal subspace there exists another,``asymmetric''
oscillatory state which has a shape of gamma-oscillations with
a phase difference of $2\pi/3$ between the slow cells and is akin to
the gamma-state described in the Sect.~\ref{sect_gamma}.

Accordingly, at sufficiently low values
of $ G_{LP}$, two qualitatively different
types of gamma-oscillations coexist: a symmetric one
and an asymmetric one in which there is a phase lag between the slow cells.
The asymmetric type, in its turn, consists of two different limit cycles:
the state in which the phase of the cell $L_1$ is $2\pi/3$ ahead of $L_2$,
as well as its mirror counterpart in which $L_2$ is ``leading''.

Depending on the initial conditions, the symmetrical network can display
either of these three gamma-patterns.
Being structurally stable, all three limit cycles survive introduction of weak
asymmetry between the connections; of course, the former symmetric state
develops in this case a small phase shift between the slow cells.
Increase of $G_{BL_1}$ at constant value of  $G_{BL_2}$ initially results
in the growth of the attraction basin of the oscillation with leading $L_1$
at the cost of the basins of two other gamma-states.
Further growth of $G_{BL_1}$ brings about saddle-node bifurcations
in which at first the former symmetric gamma-state
and then the gamma-state with leading $L_2$ disappear.

For example, at  $G_{BL_2}$=$G_{LP}$=0.03 all three gamma patterns coexist
in the range of values of $G_{BL_1}$ between 0.03 and 0.03774. The saddle-node
bifurcation of periodic orbits on the right border of this interval destroys
the former symmetric pattern. Of the two remaining states
with phase shift close to  $2\pi/3$, the pattern with leading $L_2$ disappears
in the saddle-node bifurcation at $G_{BL_1}$=0.04732: beyond this value of
$G_{BL_1}$ only one pattern of the gamma-type can be observed.

Let us return to characterization of the symmetric network with
$G_{BL_1}=G_{BL_2}$=0.03.
Except for the ``unison'' symmetric gamma-state,
the appearance and properties of the gamma- and
theta-oscillations display only small quantitative difference
from the case $G_{BL_1}\neq G_{BL_2}$ described in the preceding sections.
Two paragraphs above we have characterized
solutions which at low values of $ G_{LP}$
look like a gamma-oscillation with a phase shift between the slow cells
close to $2\pi/3$.
At around $ G_{LP}$=0.073 this pattern is rapidly transformed
(via the mixed-mode oscillation) into the theta-rhythm
in which the slow cells oscillate in the antiphase;
stable theta-oscillations persist in the whole studied range  $ G_{LP}\leq3$.
Similarly to the asymmetric case, at sufficiently high values
of conductance $ G_{LP}$,
theta/gamma-state  has not been observed, and the theta-oscillations
are the only attractor of the system (cf. Fig.~\ref{hyst}).
In absence of symmetry between $BL_{1,2}$,
the periodic theta/gamma state is eliminated in the
course of the saddle-node bifurcation. In the symmetric case, this
bifurcation is replaced by the inverse pitchfork bifurcation;
for $G_{BL_1}=G_{BL_2}$=0.03 this event takes place at $G_{LP}$=2.014.
As a result, the periodic solution of the theta/gamma type persists
at higher values of $G_{LP}$ as well, but is attracting only in the invariant
``diagonal'' subspace $u_{L_1}=u_{L_2}$ and $v_{L_1}=v_{L_2}$.
A weak violation of the symmetry in initial conditions leads to creation
and growth of the phase shift between the slow cells,
which eventually destroys the theta/gamma oscillations and replaces them
by the theta-rhythm.

Concerning the border between  attraction basins in case of hysteresis between
theta and theta/gamma rhythms, merely the  quantitative shift due to variation
of the ratio $G_{BL_1}/G_{BL_2}$ has been observed;
the shape of the basins has been largely preserved.
Compared to Fig.~\ref{initial_values_asym}(a), in the symmetric
case $G_{BL_1}=G_{BL_2}$ the attraction basin of the theta/gamma state
is slightly broader. In contrast, projection of attraction basins
for membrane variables of two slow cells
(analog of Fig.~\ref{initial_values_asym}(b),
which is symmetric with respect to the diagonal)  displays a slight narrowing
of the theta/gamma region.
Besides, we have observed that symmetry in the conductances
$G_{BL_{1,2}}$ lowers the threshold values of the membrane potentials
$v_{L_{1,2}}$, which are necessary for the onset of the theta rhythm.

\section{Discussion}
\label{sect_discuss}

In the preceding sections we have presented
a minimalistic model for
a working unit of neuronal network in the hippocampus.
Below we briefly discuss a few important aspects of this model.

\subsection{Minimality of network}

Arguably, the size of the ensemble cannot be further reduced
without introducing drastic distortions into collective dynamics.
For example, omittance of one of the slow $L$ cells would disable the theta
rhythm. As seen in Fig.\ref{switch}, in this regime two $L$ cells are
oscillating in the antiphase, and the fast pyramidal cell exhibits
two spikes within a complete oscillation cycle: one spike comes
shortly before the maximum of $L_1$ and the other one precedes
the maximum of $L_2$. As soon as one slow partner is removed,
this rhythmic pattern becomes impossible.

Further, our numerical experiments show that if the basket cell is removed
from the configuration, the rhythmic patterns persist but their dependence
on the strength of the coupling between the remaining pyramidal cell
and the slow cells is weakened. If the configuration of connections
is symmetric, the system is
restricted to only one scenario of the onset of rhythms
which is mostly determined by the choice of initial conditions.
This seems to contradict to the experimental evidence which
confirms that switching between the regimes depends on the coupling strength
\cite{Gloveli}. In the attempts to counteract the removal
of the basket cell by asymmetry in the connections {\em from} the
pyramidal cell {\em to} OLM-cells we observed that
the region of existence of the theta/gamma regime is
significantly reduced.

Thus, we suppose that presence of the basket cell allows
the neuronal network not only to ``maneuver'' between different
scenarios of emergence of all rhythms but also to expand the
region of existence of the theta/gamma regime.

The role of asymmetry in this context is twofold. On the one hand,
as shown in Sect.~\ref{sect_symmetry}, the symmetric case possesses more
attractors, which enhances the flexibility of the network. On the other
hand, reduction of variability in the strongly asymmetric network
means increase in the robustness, which can also be useful in certain
situations.

\subsection{Possible implications for larger networks}
We have started this section by defending our choice
of the minimal model. Here we turn to the ``inverse'' question: 
are patterns, generated by this minimal unit, recognizable 
on the background of large hyppocampal networks? 
Numerous experiments in the hyppocampal areas CA1 and CA3
unambiguously confirm existence of all described oscillatory states: 
the gamma, the theta and the theta/gamma~\cite{Bur, Glov,Gloveli}. 
Furthermore, in-vitro technique of directional cross-sections 
in the hyppocampus
has allowed to single out the patterns: in longitudinal
sections mostly the theta rhythm is recovered, whereas 
oscillations of gamma and theta/gamma are typical
for transverse and ``medium'' (so-called coronal) slices.
This has to do with morphology of the
participating cells. The OLM cells are arborized (possess 
dendritic structure) mostly in the longitudinal direction. 
In longitudinal slices synaptic connections from these cells
to the pyramidal and basket ones are at the
strongest; this results in the synchronous theta rhythm 
in the area CA3~\cite{Glov,Gloveli,Tor}. In other directions
arborization is much less pronounced; hence, the relative
role of the OLM cells is weaker, and one observes mostly the
gamma and theta/gamma oscillations produced by
interaction of pyramidal and basket cells.

Other kinds of cells contribute to shaping and maintenance
of rhythmic patterns as well: the study~\cite{Klaus} lists 
at least 21 classes of inhibitory interneurons which participate 
in the oscillations. Anyone of them could be a part of
the minimal module. However, high synchrony between inhibitory
interneurons responsible for gamma-rhythm~\cite {Glov, Haj} 
allows to restrict modeling to just one type:
we have chosen the basket cells, whose interactions 
with pyramidal cells are well investigated experimentally. 
On adding the OLM cells which are required for the generation of 
the theta state, we get by with just three types of cells in a module.

How do such ``modules'' interact with each other in a large-scale network? 
What cells mediate the interactions? Which type of connection --
excitatory or inhibitory -- should be considered? 
From the point of view of morphological properties, briefly sketched above, 
it seems plausible that modules communicate with each other
through OLM cells. The pyramidal and basket cells 
have been shown to be involved mostly in locally
synchronized gamma oscillations~\cite{Glov}. 
Within this conjecture, OLM cells of a module 
can inhibit pyramidal cells in the
neighboring modules~\cite{Tor}. As for interactions between OLM 
and basket cells in large networks, reliable experimental data
are still missing.
 
Spatial-temporal patterns in big networks depend on 
the strength of connection between modules.
For the case of strong coupling there is
experimental evidence of  phase theta waves~\cite{Lub}.
In contrast, moderate coupling leads to spatial-temporal 
gamma oscillations which modulate slow theta patterns (clusters)~\cite{Bel}.
Notably, such patterns are sensitive to the phase shift
between the OLM cells. Finally, if the interaction is sufficiently weak,
the modules oscillate independently of each other. In this case, 
rhythms found in a single module, persist, and presence of neighbors
results, mostly, in shifting the transitions between them. 

\subsection{Multistability}

Hysteretic properties, akin to the described effect of multistability,
have been registered in experiments with subthalamic neurons~\cite{Kass}.
In these experiments application of different bias currents
and shifting of the baseline membrane potential
resulted in the onset of such different regimes as tonic firing,
rhythmic bursts or silent upstate. According to the
experimentalists, multistability can be controlled by dynamics
of ionic channels~\cite{Kass}.

In our model, depolarization of initial membrane potential of L-cells results
in the onset of the theta/gamma regime, but stronger depolarization
switches the latter to pure theta oscillations.

To our knowledge, multistability of the discussed kind
has not yet been registered in CA3,
either in experiments or in existing theories.
(Metastable states in large sections of CA3, experimentally found
in ~\cite{Sasaki}, refer to alternating activity of different groups of neurons,
and not to different rhythmic patterns).
In particular, the authors of the more detailed model described
in~\cite{Gloveli} do not report on coexistence of attractors.
Transitions between rhythms can be related to physiological state of
a cell. The OLM cells are known to possess two specific ionic currents
responsible for the onset of theta rhythm \cite{Gloveli,Sar}.
One of these currents is activated by the strong hyperpolarization,
whereas the latter, in turn, is switched on by depolarization.
These states correspond, respectively, to the top and bottom stripes
in the left panel of Fig.~\ref{initial_values_asym},
therefore in the corresponding ranges of potentials
theta regime is established. For the initial conditions
from the ``middle'' stripe  these currents are modest, hence
theta rhythm cannot develop, and theta/gamma rhythm is observed instead.
This interpretation requires, of course, an experimental testing.

\section*{Acknowledgements}

The authors are grateful to T.~Gloveli, A.~Ponomarenko, H.~Rotstein
and S.~Schreiber for fruitful and stimulating discussions.
Research of A.L. and L.S.-G. was supported by the Bernstein Center Berlin
(Project A3); research of M.Z. was supported by the DFG Research Center
MATHEON (Project D21).

\end{document}